\documentclass{article}
\usepackage{times}
\date{}
\usepackage[T1]{fontenc}
\usepackage{amsmath}
\usepackage[cmintegrals]{newtxmath}
\usepackage{longtable}
\usepackage{pifont}
\usepackage{algpseudocode}
\usepackage{subcaption}
\usepackage{makecell}
\usepackage{float}
\usepackage{enumitem}
\usepackage[linesnumbered,ruled]{algorithm2e}
\usepackage[dvipsnames]{xcolor, colortbl}

\newcommand{\ie}{\textit{i.e., }}
\newcommand{\eg}{\textit{e.g., }}
\usepackage{hyperref}
\usepackage{cleveref}
\usepackage[sort&compress,numbers,square]{natbib}

\usepackage{tikz}
\usetikzlibrary{arrows,shapes,positioning,shadows,trees}
\usepackage{pgf-pie}
\newtheorem{mydef}{Definition}
\usepackage{soul}%
\begin{document}

\title{A Survey on the Use of Preferences for Virtual Machine Placement in Cloud Data Centers}

\author{Abdulaziz Alashaikh\and Eisa Alanazi \and Ala Al-Fuqaha\footnote{ Abdulaziz Alashaikh is with the Computer\& Networks Engineering Department, University of Jeddah, Saudi Arabia (asalashaikh@uj.edu.sa). Eisa Alanazi with Computer Science Department, Umm Al-Qura University, Saudi Arabia (eaanazi@uqu.edu.sa). Ala Al-Fuqaha is with the Computer Science Department, Western Michigan University, USA (aalfuqaha@wmich.edu).}}

\maketitle

\begin{abstract}
With the rapid development of virtualization techniques, cloud data centers allow for cost effective, flexible, and customizable deployments of applications on virtualized infrastructure. Virtual machine (VM) placement aims to assign each virtual machine to a server in the cloud environment. VM Placement is of paramount importance to the design of cloud data centers.
Typically, VM placement involves complex relations and multiple design factors as well as local policies that govern the assignment decisions. It also involves different constituents including cloud administrators and customers that might have disparate preferences while opting for a placement solution. Thus, it is often valuable to not only return an optimized solution to the VM placement problem but also a solution that reflects the given preferences of the constituents. In this paper, we provide a detailed review on the role of preferences in the recent literature on VM placement. We further discuss key challenges and identify possible research opportunities to better incorporate preferences within the context of VM placement. 
\end{abstract}

\section{Introduction}

Today, virtualization technology has facilitated the hosting of millions of services and applications on the cloud and simplified their deployments and accesses. Many small, medium, and large organizations have been moving their applications or even their entire data centers to the cloud. This is mainly motivated to their economical advantage. in terms of real-estate, equipment costs and manpower compared to private data centers. Cloud data centers also maintain high degree of flexibility where services and applications can be deployed, terminated, migrated, and replicated on-demand \cite{Pietri:2016, Sun2016}. In addition, cloud data centers offer a wide range of customizable settings and add-on features to suit the requirements and non-functional needs of different customers.

Essentially, virtualization enables dynamic sharing of physical resources such as processors, memory, and storage of the cloud infrastructure. 
An application is installed on a logically isolated Virtual Machine (VM) that has limited and pre-configured access to the shared resources of Physical Machines (PM) and networking devices. 
The process of mapping the VMs to the existing PMs is commonly known as the Virtual Machine Placement (VMP) problem. 
The VMP problem is of paramount importance to the design and operation of cloud data centers. Throughout the last decade, there have been many works addressing the VMP from different perspectives. 
The majority of these works are concerned about the profitability of data centers and the energy bill as well as other operational aspects. 
While many proposals share similar objectives, they may have taken different approaches, yielding a wide range of frameworks, models, and algorithms.

A pre-requisite to the success of existing approaches is handling customer  and administrator preferences when looking for a placement solution. This is due to the fact that cloud administrators and customers are often not satisfied with \emph{any} placement solution. Rather, they are interested in the best, or at least good, solution that meets their preferences. As a motivating example, consider Bob who wants to place a VM that requires one unit of CPU and two units of RAM. Alice, on the other hand, has a VM that requires two units of CPU and one unit of RAM. The cloud data center, which is administrated by Charlie, has two physical machines $P_1$ with 3 CPU units and 4 RAM units and $P_2$ with 4 CPU units and 3 RAM units. Assume the objective is to minimize the total wastage of RAM and CPU units. Obviously, the four possible placements $(P_1,P_1)$,$(P_2,P_2)$,$(P_1,P_2)$, and $(P_2,P_1)$ are all Pareto optimal with total wastage of 8. 

Now, assume both Alice and Bob prefer placing their VMs in a PM with faster CPU. In such case,the solution $(P_2,P_2)$ is preferred to all other solutions. Furthermore, in case this solution is unattainable (\eg violating a constraint), $(P_1,P_2)$ and $(P_2,P_1)$ are both preferred to $(P_1,P_2)$. These placements represent the most preferred feasible options given both Bob and Alice preferences. Finally, Charlie prefers Bob's VM be placed on $P_1$ to better meet Bob's service level agreement (SLA) and therefore, $(P_1,P_2)$ is the optimal placement given the three preferences.  This example demonstrates that cloud customers and administrators may have preferences that need to be considered while consolidating customer VMs (\eg customers prefer servers with faster CPUs) and solutions are no longer indifferent but some are preferred to others.

There has been a number of attempts trying to capture preferences during the VMP process. 
For instance, statements of the form ``I prefer $X$ and $Y$ to be placed on a different geographic location", ``I prefer $X$ not to be co-located with other VMs", or "I prefer $X$ and $Y$ to be placed on the same PM" are particularly popular in the literature as customer preferences \cite{Ye2015,Tsakalozos2011,Tsakalozos2013,Espling2016,Deshpande2017,Dow2016,Alashaikh2019}. 

Furthermore, the cloud administrator is usually overwhelmed with a large set of potential placements and has several preferences on the deployment process. While any solution in the set is optimal to the underlying constraints and objectives, some solutions are expected to be preferred (by the administrator or customer) compared to others \cite{jung2010performance, jayasinghe2011improving, wang2007appliance, van2009autonomic, Tsakalozos2013}. 
Neglecting the preferences while finding a solution would discourage the customers from subscribing to the cloud and administrators from using the proposed approach. Hence, proper handling of preferences is a critical issue to address in the design of a cloud. 

The relevant literature is rich with survey papers on the VMP problem. There are multiple VMP survey papers that interested readers may refer to, such as \cite{Whaiduzzaman2014, Mann2015, Lopez-Pires2015, Pietri:2016, Sun2016, Masdari:2016, SilvaFilho2018, MohamadiBahramAbadi2018, Attaoui2018, Zhang2018,Talebian2019}.
Most of these surveys provide multiple classifications and reviews of approaches to handle the VMP problem including modeling, the objectives and constraints representation and formulation, the techniques used to solve the problems, the scope and the context in which the placement is taking place and more profoundly to its variants. 
The focus on this paper is different as it aims to look deeper into the notion of preferences in the VMP context. We begin with a clarification of what is meant by preferences in the context of an  optimization problem. \Cref{fig:pyramid} shows a conceptual difference between the three types of knowledge. Basically, the objectives and constraints of the problem instantiate its optimality and feasibility regions, whereas satisfaction is commonly attributed to preferences \cite{Sonnek2010}. Indeed, one can promote a  preference to an objective or constraint. Many attempts in the literature adopt this methodology and add preferences as an objective to the problem. We believe such transformation albeit perhaps efficient, has its own drawbacks as it is not always possible to construct objectives that reflect the preferences without losing some information. Especially when the customer or administrator pose complex preference statements over different aspects of the network. 

\begin{figure}[]
	\centering
	\begin{tikzpicture}[scale=.7]
	\small
	\fill[gray!55!red!35!, path fading=north] (0,0) -- (8,0) -- (4,6) --(0,0);
	\draw[line width=.75pt] (0,0) -- (8,0) -- (4,6) --(0,0);
	\draw[] (2.5,3.75) -- (5.5,3.75);
	\draw[] (1.2,1.8) -- (6.8,1.8);
	\node[] at (4,.8) {Constraints};
	\node[] at (4,2.5) {Objectives};
	\node[] at (4,4.25) {Preferences};
	\draw[->, >=latex, orange!60!gray!30!, line width=15pt]   (-2,5.5) to node[black]{\rotatebox{90}{ \bf \qquad    Must be satisfied}} (-2,0);
	\end{tikzpicture}
	\caption{The satisfaction level expected for different types of information:  Constraints are requirements that every solution needs to satisfy; Objectives are the deriving force in searching and choosing one feasible solution over another in the feasible region; Preferences represent wishes and desires. Objectives and constraints are part of the problem representation and thus need to be taken into account in any algorithm while preferences are the key concept in personalizing an approach to the customers and administrators.}
	\label{fig:pyramid}
\end{figure}
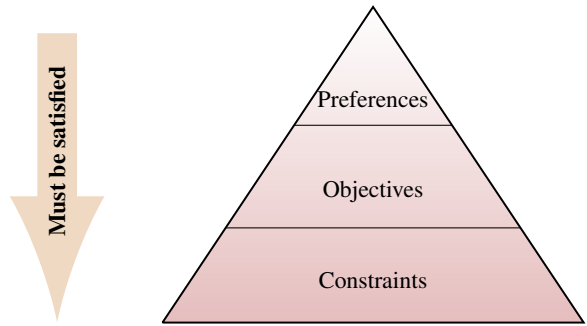
\begin{figure*}[!h]
	\centering
	\small
	\tikzset{
		every node/.style={draw,text width=2cm,drop shadow},
		style1/.style= {rectangle, rounded corners=2pt, thin,align=center,fill=gray!55!red!35!,text width=4cm},
		style2/.style= {rectangle, rounded corners=6pt, thin,align=center,fill=orange!60!gray!20!white},
		style3/.style= {rectangle,thin,align=left,fill=gray!15}}
	\begin{tikzpicture}[scale=.65,remember picture,level 1/.style={sibling distance=65mm},edge from parent/.style={->,draw},>=latex, level distance=1.75cm]
	\node[style1] {VMP with Preferences}
	child {node[style2] (c1) {Source}}
	child {node[style2] (c2) {Type}}
	child {node[style2] (c3) {Approach}}
	child {node[style2] (c4) {Problem}};
	\node [style3,below of = c1,xshift=15pt] (c11) {DM};
	\node [style3,below of = c11,xshift=15pt] (c111) {\small Customer};
	\node [style3,below of = c111] (c112) {\small Cloud Admin};
	\node [style3,below of = c112] (c12) {VM};
	\node [style3,below of = c12] (c13) {PM};
	\node [style3,below of = c2,xshift=15pt] (c21) {Utility};
	\node [style3,below of = c21] (c22) {Importance};
	\node [style3,below of = c22] (c23) {Binary Relation};
	\node [style3,below of = c23] (c24) {Soft\\ Constraints};
	\node [style3,below of = c3,xshift=15pt] (c31) {Metaheuristics};
	\node [style3,below of = c31] (c32) {Matching};
	\node [style3,below of = c32] (c33) {MCDM};
	\node [style3,below of = c33] (c34) {Heuristics};
	\node [style3,below of = c4,xshift=15pt] (c41) {Placement};
	\node [style3,below of = c41] (c42) {Re-Placement};
	\node [style3,below of = c42] (c43) {Migration};
	\foreach \value in {1,2,3}
	\draw[->] (c1.195) |- (c1\value.west);
	\foreach \value in {1,2}
	\draw[->] (c11.195) |- (c11\value.west);
	\foreach \value in {1,2,3,4}
	\draw[->] (c2.195) |- (c2\value.west);
	\foreach \value in {1,...,4}
	\draw[->] (c3.195) |- (c3\value.west);
	\foreach \value in {1,2,3}
	\draw[->] (c4.195) |- (c4\value.west);
	\end{tikzpicture}
	\caption{General Classification of preferences based approaches on Virtualization}
	\label{fig:classification2}
\end{figure*}
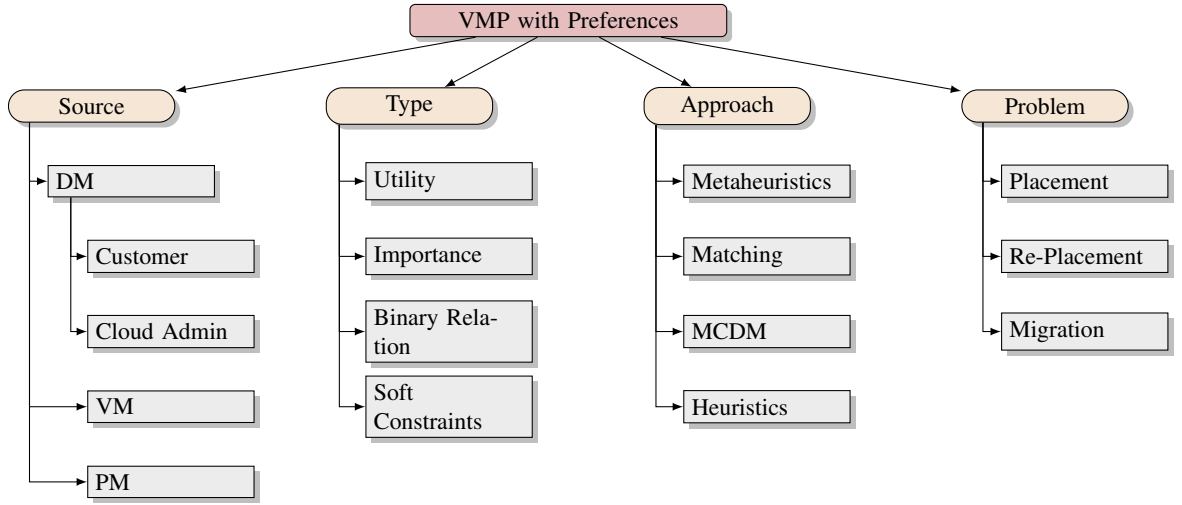

In this paper, we provide a detailed survey of VMP proposals that incorporate soft constraints and preferences highlighting their types, implications, and the algorithms used to handle them. We also discuss some challenges and identify possible research opportunities to better incorporate preferences in the context of VMP. A classification of the proposed approaches is given in \Cref{fig:classification2}. Some of the important aspects that need to be addressed when proposing an approach include: 
\begin{description}

\item[Source] identifies who initiates the preferences, the customer who owns the VMs, the cloud administrator who owns the PMs, or both. 
\item[Type] depicts how the preferences are represented. (\ie via utility function, binary relation, etc.) 

\item[Approach] classifies the algorithm family used to develop an algorithm. In a broad sense, they can be classified to heuristic, metaheuristic, and exact algorithms. 
\item[Problem] describes the VMP problem context as there are different variations of the problem, such as migration and re-placement.
\end{description}
{It is important to realize that the different considerations of these aspects impose some limitations on how the preferences are represented and processed. 
More details about these aspects and their implications will be discussed throughout the paper.}

The rest of the paper is organized as follows: 
\Cref{sec:background} presents background regarding the VMP problem and preferences. 
We then review the recent literature on the VMP problem with preferences in \Cref{sec:algorithms}. 
\Cref{sec:benefits} provides our insights on the potential benefits for the use of preferences in the context of the VMP problem, followed by a discussion over current limitations and future research opportunities. 
\Cref{sec:conclusions} concludes this work.

\section{Background}
\label{sec:background}
\subsection{VMP design problem}

\begin{figure*}[t]
\centering
\includegraphics[width=.85\textwidth]{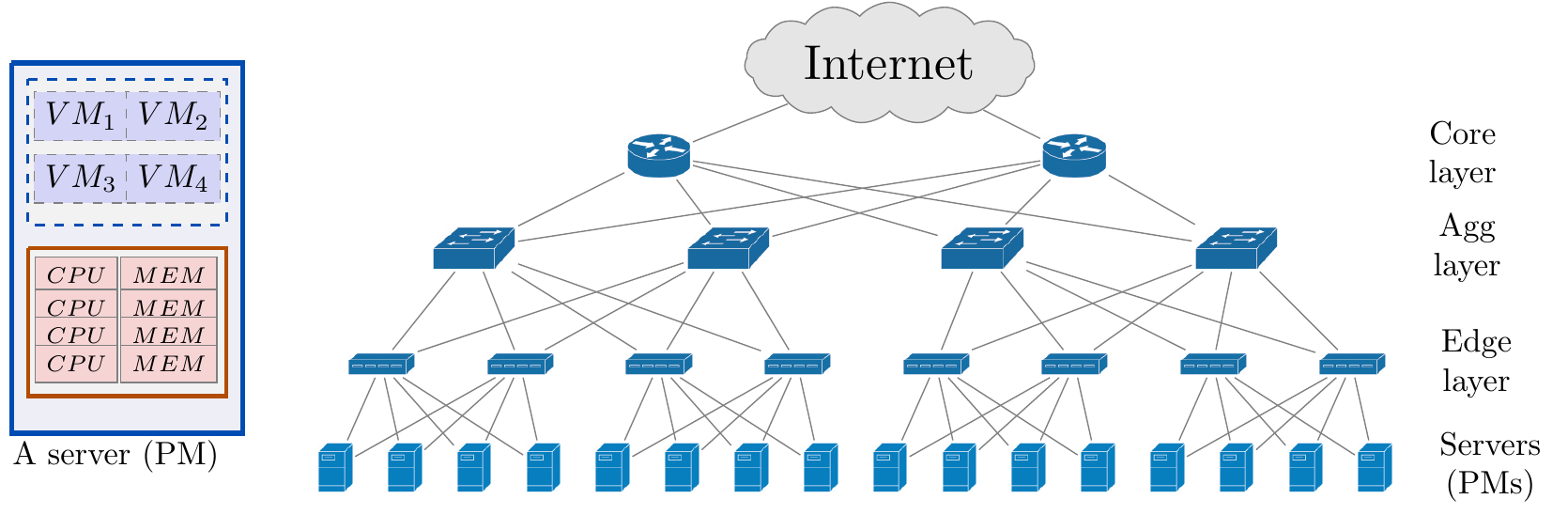}
\caption{A 3-layer tree topology of a data center.} \label{fig:data center}
\end{figure*}
VMP is an essential process for the design and operation of cloud data centers.  
VMP is concerned with the assignment of virtual machines to physical machines (servers) in the data center. 
The decision of such placement highly impacts the performance of the data center in terms of
utilization levels, traffic congestion level, and energy consumption, all of which the Decision Makers (DMs) continuously try to optimize. 

A wide range for data center network topologies have been studied extensively in the literature. 
Basically, the architecture tells us how routers, switches, and servers are connected, and what is the number of ports required on these devices to implement the network. 
The choice of a topology has a great impact on the cost, scalability, resiliency (in terms of alternative paths), complexity in terms of number wires, and bandwidth and traffic distribution within the data center. 
As an example, consider the tree topology shown in \Cref{fig:data center}. It is comprised of three layers; namely, Core, Aggregation, and Edge.  The core layer is the data center gateway to the public Internet, whereas the switches in the edge layer interconnect the servers (called physical machines) to each other and to the gateway through the middle layer switches. 
This architecture is a classic one that is commonly used as a reference or baseline for evaluating other proposals. It is widely adopted due to its simplicity. 
Despite that, it raises a number of technical concerns regarding its scalability and resiliency. 
Hence, modern data centers have evolved to utilize more sophisticated architectures such as Clos tree, recursive and random topologies. 
More about alternative architectures and their feature can be found in \cite{Qi2014,Yao2014, Colman-meixner2016, Shooshtarian2018}. 

Many virtualization tool developers have realized the importance of the customizability of the virtual services and also enabled VMP based on multiple attributes that also consider non-technical ones,  \eg Profile-Driven Storage in vSphere 5.0 \cite{vmware1,vmware2}, HP Moab \cite{hpmoab}, Server affinity policies in Oracle VM server \cite{oracle}, and many others \cite{AmazonWebServices2018}.  

In realization to consumers varying needs, cloud service providers have designed multiple options for VMs owners to choose from, in which an owner can choose from a set of pre-defined VM types (\ie a combination of disk space, allocated memory, bandwidth, etc.) \cite{googlemachinetypes, ec2}. On top of that, many providers went to offer added-value or policy-based options, such as colocation, geo-redundancy, availability, etc (see for instance, placement policies and proximity placement group of Microsoft Azure \cite{azure1,azure2},  phoenixNAP colocation and security \cite{phoenixnap}, 
IBM High availability \cite{ibmHA}
, and most recent mobile core network virtualization on AWS \cite{awspco}).

A basic VMP problem can be defined as follows: Given $\mathcal{PM} = \{P_1, P_2,..., P_m \}$, a set of $m$ physical machines ({$P_k$}) and $\mathcal{VM} = \{V_1, V_2,..., V_n \}$, a set of $n$ virtual machines ($V_i$), how to place each virtual machine into a physical machine while optimizing a set of objectives and satisfying a set of constraints? 

\subsubsection{Objectives}
Basically, the utmost goal when solving a VMP problem is to optimize some metrics of interest. These metrics are identified as objectives. From a cloud administrator's perspective, the main objectives considered in the VMP are highly related to maintaining profitable operation of the cloud. 
On top of the list of these objectives is energy consumption. 
Typically, energy consumption is modeled in terms of CPU utilization, and a possible minimization approach is to distribute the load among existing PMs to reduce CPU usage on each PM. Alternatively, turning off underutilized PMs and move their VMs to other PMs appeared to be more a effective strategy. However, this approach brings the issue of wear-and-tear to the surface as PMs maintaining higher resource utilization are expected to fail sooner than others. 
In addition to energy consumption, efficient packing of VMs with low defragmentation (resource wastage index) increases the ability to accommodate more VMs in the cloud which is an important feature from the profitability perspective. 
Moreover, VMs may need to communicate with each other to perform a certain task. 
If two communicating VMs are hosted on the same PM, then no traffic will go through the switches. 
However, if they reside on two different machines, traffic has to go through intermediate switches. This kind of traffic is called intra-data center traffic as opposed to external traffic from/to the Internet. 
Excessive intra-data center traffic may disrupt traffic between VMs and Internet as intermediate switches may become congested. 
In order to prevent this from happening, the traffic between peer VMs needs to be localized as much as possible. This can be achieved by placing VMs that are highly interdependent close to each other. 
However, this is mainly constrained by the capacities of the PMs. 

The objectives may also involve other metrics that contribute to the profitability index. 
For example, traffic minimization to avoid unpredicted/undesired congestion that may degrade performance, QoS satisfaction, infrastructure reliability, VM availability, energy consumption of switches, routers, and auxiliary equipment (\eg cooling systems), etc.
For more examples on alternative objectives the reader is referred to \cite{Lopez-Pires2015,Mann2015,Attaoui2018, SilvaFilho2018, MohamadiBahramAbadi2018, Zhang2018}. 

Generally, a single objective or multiple objectives can be involved in making the placement decision. However, due to the nature of the VMP problem, involving multiple dimensions, factors, stakeholders, and influential metrics, researchers usually consider multiple objectives. 
As discussed above, these objectives are highly interdependent and likely conflicting with each other. 

\subsubsection{Constraints}

Mostly, any VMP problem can be seen as a pin packing problem. 
Hence, the constraints are more or less similar to the pin packing problem: the assignment constraint ensures that each VM is assigned to one PM, and one PM can host multiple VMs, and the capacity constraint ensures the availability of enough resources on the PM to utilize and fulfill VM requirements. 
Typically, each VM requires resources described in terms of CPU cycles, RAM space, disk space, network bandwidth, etc. 
While a server $P_k$, leveraging virtualization techniques, can host more than a single VM, the resource utilization (\ie load) should not exceed its resource capacities. 
In addition, other design considerations can be imposed as constraints. 
For example, a cloud administrator may be interested in setting a limit on the number of VMs migrations to avoid instability and interruption of the cloud operation, or impose a constraint on security levels for some VMs.

\subsubsection{Variants}

%
%
There are multiple variants of the basic VMP problem that appeared in the literature. 
The basic VMP problem may assume that there are some VMs hosted in a DC and new VMs are to be placed on PMs based on the current utilization and operational status of the DC. 
On the long-run, this might lead to defragmentation of the resources. Therefore, rearrangement of the VM-PM mapping is desired in which multiple VMs are consolidated into a PM to enhance the mapping efficiency \cite{SunilRao2015}. 
Migration is the technique used to move a running VM from its current PM and re-assign it to a new PM. It is a powerful technique that enables the cloud operator to reallocate resources and reconfigure assignment based on current status of the machines \cite{Zhang2018}. 
However, there are critical considerations when making the migration decisions, particularly, related to the timing, frequency, and the cost of migrations \cite{Verma:2010}. 
Hence, these factors exacerbate the complexity of problem in which adopting soft rules or preferences to handle them and introduce a meaningful trade-off is advisable. 
In our study, we refer to these VMP problem variants when classifying the research efforts that appeared in the literature.

VMP can also be classified based on the number of the clouds involved \cite{Mann2015}. 
A single-cloud (private or public) VMP problem is concerned with the placement of VMs into PMs within the same data center. 
Multiple clouds VMP, typically, entails solving two assignment problems in which the system has to distribute incoming VM requests over the different clouds and then optimize each cloud individually. 
There are multiple architectures and models for multi-cloud VMP. For example, hybrid cloud \cite{Unuvar2014}, 
multiple cloud with multiple administrators \cite{Saber2018}, and others \cite{Menzel2012, Grozev2014, Jamshidi2017, Jain2018}. Another related but different problem is the cloud administrators selection problem where it is assumed that customers express preferences and intermediate brokers negotiate for the best offer \cite{Amato2013, Kella2014, Mahalingam2015, Mezni2018}.

\subsection{Preferences}
\label{sec:pvmp}
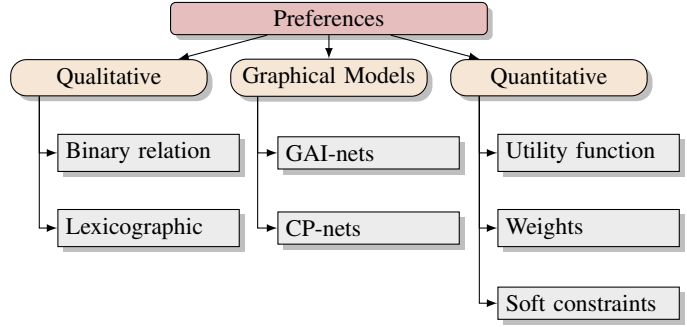
\begin{figure}[t]
	\centering
	\small
	\tikzset{
		every node/.style={draw,drop shadow},
		style1/.style= {rectangle, text width=2cm, rounded corners=2pt, thin,align=center,fill=gray!55!red!35!,text width=4cm},
		style2/.style= {rectangle, text width=2.4cm, rounded corners=6pt, thin,align=center,fill=orange!60!gray!20!white},
		style3/.style= {rectangle,text width=2.2cm, thin,align=left,fill=gray!15}}
	
	\begin{tikzpicture}[scale=.45,remember picture,level 1/.style={sibling distance=65mm},edge from parent/.style={->,draw},>=latex, level distance=1.75cm]
	\node[style1] {Preferences}
	child {node[style2] (c1) {Qualitative}}
	child {node[style2] (c2) {Graphical Models}}
	child {node[style2] (c3) {Quantitative}};
	\node [style3,below of = c1,xshift=15pt] (c11) {Binary relation};
	\node [style3,below of = c11] (c12) {Lexicographic};
	\node [style3,below of = c2,xshift=15pt] (c21) {GAI-nets};
	\node [style3,below of = c21] (c22) {CP-nets};
	\node [style3,below of = c3,xshift=15pt] (c31) {Utility function};
	\node [style3,below of = c31] (c32) {Weights};
	\node [style3,below of = c32] (c33) {Soft constraints};
	\foreach \value in {1,2}
	\draw[->] (c1.195) |- (c1\value.west);
	\foreach \value in {1,2}
	\draw[->] (c2.195) |- (c2\value.west);
	\foreach \value in {1,2,3}
	\draw[->] (c3.195) |- (c3\value.west);
	\end{tikzpicture}
	\caption{A General classification of Preferences from representation point of view.}
	\label{fig:classification1}
\end{figure}

Preferences are ubiquitous. Whether we want to choose a restaurant for lunch, watch a movie or purchase an airline ticket, we often have preferences over the set of available alternatives. Thus, preferences play a key role in the decision making process and have been an active topic of research in many fields such as artificial intelligence, economics, and operations research \cite{keeney1993decisions}. Preferences can be broadly classified based on their representation as quantitative and qualitative. In quantitative preferences, users express preferences with numerical values representing the utility of having one alternative. On the other hand, qualitative preferences are expressed based on comparative assessments of  alternatives. Figure \ref{fig:classification1} shows a general classification of preferences. The graphical models refer to a class of models that have been introduced lately in the literature to represent the preferences in a compact way. They rely on the notion of (conditional) preferential independency  among attributes and can be described in terms of vertices and edges of a graph \cite{amor2016graphical}. In what follows, we review the basic methods used to represent preferences  (interested readers can refer to \cite{fishburn1970utility} for a comprehensive introduction to the field). Informally, preferences are any additional knowledge that is not part of the problem requirements and objectives, but represent the subjective judgments of the users in terms of liking or disliking an alternative. 

 In decision theory, preferences provide a way to evaluate alternatives based on subjective judgments. It is typically assumed that every alternative $\bf x$ is a vector over some $k$ attributes ${\bf x}=(x_1,x_2,\dots,x_k)$ that represent the world. For instance, in the context of VMP, an alternative can be represented by a possible placement of the VMs to PMs and the set of alternatives represent the placements that are possible given the problem constraints and objectives. Given a set of alternatives $S$, there are two important questions that need to be answered by the preference model:
\begin{itemize}
	\item (optimality) What is the best alternative given the preferences?
	\item (dominance) Given two possible alternatives $\bf x$ and $\bf y$, which one is more preferred? 
\end{itemize}

One of the simplest models for preferences is the weighted sum model obtained by assigning an importance weight $w_i$ to every attribute $i$. Then alternatives are assessed based on their value \ie $w({\bf x})=\sum_{i=1}^{k}w_ix_i$ where $w_i\geq 0$ and $\sum_{i=1}^{k} w_i=1$. For instance, consider three possible alternatives $x_1=(10,20,30)$, $x_2=(30,10,20)$, and $x_3=(20,5,25)$ defined on a set of three attributes $a_1,a_2,a_3$ with importance weights $0.6$ for $a_1$ and $0.2$ for both $a_2$ and $a_3$. Using the weighted sum model, $w(x_1)=0.6\times 10 + 0.2\times 20 + 0.2\times 30=16$, similarly $w(x_2)=24$ and $w(x_3)=18$. Given these values, one can conclude that $x_2$ is the best alternative, as it has the highest value among other alternatives and also $x_3$ is preferred $x_1$ since its value is higher. 

Another representation is the value (or utility) function where alternatives are mapped to real numbers $\mu:S\rightarrow \mathbb{R}$. Here, every alternative $\bf x$ is mapped to a real number $\mu({\bf x})$ that represents the desirability of the user for such alternative. Considering the above example in the weighted sum model, we can have $S=\{x_1,x_2,x_3\}$ and the alternative utility is simply its sum, \ie $\mu(x_1)=16$, $\mu(x_2)=24$, and $\mu(x_3)=18$. In general, an arbitrary numerical value can be associated with every possible alternative expressing its preference to the decision maker. 

Generally, a preference over a set of alternatives $S$, can be defined as a binary relation on $S$ \cite{keeney1993decisions}. Several properties arise to different relations. We start by general definition of preference relations.

\begin{mydef}
A preference relation $\succeq$ is a preorder if $\succeq$ is reflexive and transitive. 
\end{mydef} 
A preorder is called a partial order if it is also antisymmetric. A statement of the form $x\succeq x'$ is interpreted as $x$ is at least as good as $x'$. Given two alternatives $x,x'$, one can derive some useful conclusions between the two as follows
\begin{enumerate}
\item $x$ is strictly preferred to $x'$ iff $x\succeq x'$ and $x'\not\succeq x$.
\item $x$ is indifferent from $x'$ iff $x\succeq x'$ and $x'\succeq x$.
\item $x$ is incomparable with $x'$ iff $x\not\succeq x'$ and $x'\not\succeq x$.  
\end{enumerate}
Incomparability captures the scenario where there is no possible information in $\succeq$ to assess the two alternatives. However, being indifferent means the user actually prefers both alternatives to the same level. One important property that is usually assumed in practice is the completeness of the partial order.

\begin{mydef}
	A partial order $\succeq $ is said to be complete iff $x\succeq x'$ or $x'\succeq x$ for any $x,x'\in S$.
\end{mydef}
A complete partial order is also known as a linear order or a ranking over a set $S$. In such case, any two alternatives are comparable and one is preferred to the other. Lexicographic preference relations are also common in practice where an importance order over attributes are assumed.

 Another model of preferences that has gained much attention in recent years is the soft constraint model. In such model, constraints are associated with numerical values representing the cost, importance, or preference and unlike hard constraints, alternatives are not required to satisfy all of them. A formal representation for soft constraints is Valued Constraint Satisfaction Problem (VCSP) \cite{schiex1995valued}. Lastly, the concept of reference point is widely used in the evolutionary algorithms to capture users' idealized vision of the solution \cite{deb2006reference} and define a distance to relatively evaluate the closeness of a given alternative to the reference point.

\subsection{Applications of preferences in communications problems}
Given the rise of multiple aspects involved in the design of communications system that are of interest to decision makers (\eg delay, availability, utilization level, hop-count etc.) 
as well as concepts or design features that are desired to achieve (\eg load balancing, fairness, human-centric design, etc.), 
many communications problems have been regarded today as multi-dimensional or multi-criteria problems. 
Hence, incorporating the preference or soft constraint concept into the problem can be useful to many applications in communications beyond the VMP problem. 

There are many problems in the communications and networking field where it is advisable to consider such preferences and/or soft constraints. 
Such problems include multiple decision makers in which each entity has its own requirements (solid constraints) and preferences, or a problem where multiple metrics are involved in the design with some sort of priority or importance (\eg delay is more important than reliability) and require insightful handling of preferences in the design \cite{Climaco2016}. 

Consider for example, a multi-criteria routing problem. Typically, one metric is regarded as an objective while others are given as constraints, or alternatively the multiple metrics are aggregated into a single objective function with some defined weights. 
In a multi-criteria approach, the problem is approached from a different perspective where metrics to be optimized are handled as objectives. This promotes optimization in all dimensions of the problem which results in multiple solutions scattered in hybrid-planes. 
Using preferences (\eg bounds, sequences, etc.), solution space can be partitioned into smaller regions each has a subset of the solutions with different desirability level \cite{Climaco2019}. 

Similarly, a telecom network planning with multiple cost types (\eg investment, maintenance, management, extension, etc.) and different design criteria (\eg capacities, availabilities, diameter, edge degree, etc.) is another example where the adoption of preferences can help achieve more desirable outcome \cite{Collette2004}. 
This also can be extended to multilayer network design to perform mapping between different layers or design overlay networks \cite{Risso2018}, traffic grooming \cite{Rubio-Largo2013}, 
or resource provisioning considering the cost of each layer independently \cite{Xing2019}. 

Another example in communication networks is the design of wireless sensor networks which entails multiple design sub-problems such as sensor placement, data aggregation, power assignment, etc. Typically, each considers a set of objectives to be optimized (\eg detection probability, latency, deployment cost), and a proper definition of objective preferences or layout preferences would improve our perception to the design 
\cite{Konstantinidis2010,Rajagopalan2011,Murugeswari2016}.

Furthermore, the rapid evolution in many technological fields in communications as well as the complexity associated with design decisions stems from the nature of these problems, being multi-criteria, multi-attribute, multi-stakeholders, 
make the adoption of the preference concept a more viable and promising option. 
There are many problems where the application of the preference concept can be seen advantageous in modeling and solving the problem. 
For example, in the multiobjective SDN controllers placement problem, trade-offs between objective metrics are needed \cite{Sahoo2018} \cite{Jalili2019}. 
Another example is the resource scheduling problem in mobile networks where multiple users of different requirements and classes of service are being served with limited radio resources and changing conditions. 

\begin{figure}[t]
	\centering
	\begin{tikzpicture}[scale=.7]
	\footnotesize
	\pie[text=pin, color = {gray!20, gray!40, gray!60}]
	{51.0/Functional, 30.6/Non-Functional, 18.4/Mixed}
	\end{tikzpicture}
	\caption{The distribution of the preference types in the literature from functional and non-functional perspectives. 51\% of the proposed approaches deal only with functional preferences neglecting users' genuine preferences on the problem.}
	\label{fig:pktypes}
\end{figure}
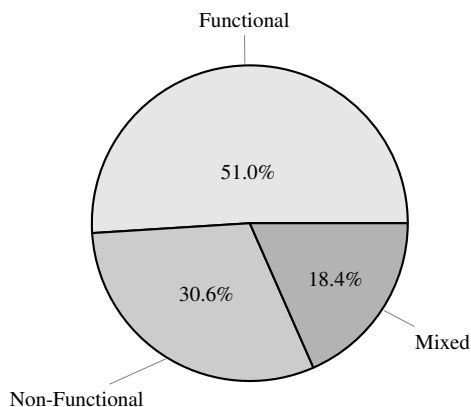
\section{Literature Survey}
\label{sec:algorithms}

\citet{Pietri:2016} surveyed the literature addressing the VMP problem, and provided a set of common optimization techniques used to achieve the desired objectives. 
	In this section, we survey existing techniques that incorporate the notion of preference and its variants in the context of the VMP problem. 

In the context of the VMP problem, preferences have been tackled by means of hints, soft constraints, soft SLAs, relaxation bounds, weights on the objectives, weights on the constraints, and affinity. One natural classification is to distinguish between preferences that are issued directly by the user (\ie cloud administrator or customer) and preferences that are provided implicitly from a certain measurement (\eg ranking VMs based on current communication cost). The later are related to measurable system attributes. For example, CPU usage index, memory usage index, and inter-VM traffic are operational factors that can be translated into preference relations or rankings. A commonly proposed approach of this type is the affinity-based VMP problem. It adopts the notion of affinity to evaluate which VM is preferred to be hosted on a PM, and which PM a VM prefers \cite{Chen2010,Sonnek2010}.  

The former preferences are derived from people interacting with the system including cloud administrators and customers. For example, a cloud admin may prefer to assign some VMs to newer PMs due to cost considerations, or a customer prefers to have his VMs located on the same PM or without sharing other VMs \cite{Tsakalozos2013, Rane2015,Alashaikh2019}. \Cref{fig:pktypes} shows the distribution of the approaches from functional and non-functional perspectives. 

 \Cref{tab:prefUsage} lists some examples of preference uses reported in the recent literature. Clearly, the literature shows a wide variety of applications. For instance, preferences have been utilized to maintain higher degree of non-functional aspects such as SLA satisfaction and availability. Also, they have utilized to represent the desire to assign sub-systems to servers with higher resources, or to servers in certain geographic locations. One notable application is ranking PMs based on their available resources and ranking VMs based on their required resources. Such ranking is often used in the matching algorithms. 
\begin{table*}
	\centering
	\footnotesize
	 \renewcommand{\arraystretch}{.5} 

		\label{tab:prefUsage}
		\vspace{-2cm}
			\begin{tabular}{|m{.12\textwidth}|m{0.16\textwidth}|m{0.6\textwidth}|}
			\hline
			Preference & References & Usage example \\ \hline
			Resources & \cite{Arianyan2016, Azougaghe2017, Chen2010, Chu2017, CPham2017, Cui2015, Cui2016, Georgiou2013, Kabir2015, Kella2014, Korasidis2017, Korupolu2009, LiWeiLing2017, Tarighi2010, Tarighi2010a, Wangj2016, Xu2013, Xu2015, Amato2013} & \begin{itemize} \item A VM ranks PMs based on available resources or resources utilization ratio. \item A PM ranks VMs based on required resource/s. \item A VM prefers a PM with faster CPU. \end{itemize} \\ \hline
			Location & \cite{Li2016, Pillai2016, Sotiriadis2016, Tantawi2016} & \begin{itemize} \item A customer prefers a geographical location over another for security or reliability concerns. \item A cloud administrator prefers to place a group of VMs in a specific location for organizational purposes. \end{itemize} \\ \hline
			Inter-VM dependency & \cite{CPham2017, Liu2017, Liu2018, Sonnek2010, Xu2015,Ranjana2019} & \begin{itemize} \item Preference information is derived from expected/measured traffic between VMs. \item VMs collaborating in performing specific tasks or services. \end{itemize} \\ \hline
			Colocation & \cite{Chen2017, Dhillon2013, Dow2016, Sonnek2010, Tantawi2016, Unuvar2014, Espling2016} & \begin{itemize} \item VMs belonging to the same customer are preferred to be placed close to each other. \item Flexible split preferences \ie allow group VMs to be placed on different clouds or in the same cloud. 
			\end{itemize} \\ \hline
			Anti-colocation & \cite{Chen2017, Dow2016, Georgiou2013} & \begin{itemize} \item Place VMs disjointly for security, reliability, or administrative purposes. \end{itemize} \\ \hline
			Sharability & \cite{Mahalingam2015, SunilRao2015} & \begin{itemize} \item A customer prefers to not share a PM with others. \end{itemize} \\ \hline
			Migration cost/overhead & \cite{Kella2014, LiWeiLing2017, Xu2011, Zhou2013} & \begin{itemize} \item VMs with lower migration overhead are more preferred to migrate. \item VMs with migration cost (\eg susceptible to SLA violation due to migration) are less preferred to migrate. \end{itemize} \\ \hline
			Availability/ Reliability & \cite{Zhou2013, Araujo2018} & \begin{itemize} \item Include the degree of PM wear-and-tear (\eg due to frequent PM ON/OFF cycles) in weighted function. \item Replicate a VM to improve its availability. \end{itemize} \\ \hline
			Cost & \cite{Azougaghe2017, Araujo2018, Jain2018} & \begin{itemize} \item A customer may prefer a lower cost PM. \item A cloud admin may prefer to place VMs based on their payment scale and the cost of the infrastructure. \item Typically, the cost computed as utility function composing multiple attributes or just monetary value.\end{itemize} \\ \hline
			Ordering & \cite{Alashaikh2019, He2011, Ihara2015} & \begin{itemize} \item Preference order of preferred PM to host a VM or a set of VM. \end{itemize} \\ \hline
			Distance & \cite{Chu2017, Kim2014, Xu2011} & \begin{itemize} \item An application prefers computing and storage PMs pair with smaller distance. 
				\item An application prefers to be placed near a specific PM. \end{itemize} \\ \hline
			Objectives & \cite{Addya2017, Blagodurov2015, Deng2014, Lopez-Pires2017, Saber2018, Viswanathan2011, Ihara2015} & \begin{itemize} \item The extent to which an objective is prioritized over other objectives. \end{itemize} \\ \hline
			SLA & \cite{Shi2011, Zhou2013} & \begin{itemize} \item Soft SLA tolerates some violations.  \end{itemize} \\ \hline
			Security & \cite{Azougaghe2017, ZhangY2017} & \begin{itemize} \item Security preferences can be modeled as a function of the system diversity. 
				\item Place VMs according to their security preferences and PMs capabilities. \end{itemize} \\ \hline
			Other/ Unspecified & \cite{Rane2015, Tsakalozos2011, Tsakalozos2013, Wright2012, Ooi2012, Li2016, Sedaghat2013,Feng2019} & \begin{itemize} \item a generic model in which preferences information are customized/defined by the user/DM. \item Place a VM in a specific PM \cite{Li2016,Sedaghat2013}. \end{itemize} \\ \hline
		\end{tabular}
	\caption{Existing application scenarios for preferences}
\end{table*}

In the remainder of this section, we review the literature based on the proposed approaches. We classify existing works into five broad classes of algorithms: Metaheuristics, Matching, MCDM, Heuristics, and others. \Cref{tab:methods} shows an overview of the existing works based on this classification. At the end of this section, we compararatively discuss the advantages and shortcomings of the five classes. 

\begin{table*}
	\scriptsize
\centering
	\caption{Existing Approaches that handle preferences in VMP}
	\label{tab:methods}
		\begin{tabular}{|m{3cm}|m{8cm}|}
			\hline
			\textbf{Method} & \textbf{References}  \\ \hline
			Metaheuristics & Simulated annealing \cite{Tsakalozos2011, Tsakalozos2013, Addya2017},  NSGA-II \cite{Blagodurov2015, Alashaikh2019}, Improved grouping genetic algorithm \cite{Deng2014}, Memetic Algorithm \cite{Ihara2015, Lopez-Pires2017} Ant-colony \cite{Liu2017, Liu2018}, heuristics \cite{Mahalingam2015} Hybrid algorithm \cite{Saber2018}   \\ \hline
			Matching & Stable matching \cite{Korupolu2009, Dhillon2013, Xu2013, Kim2014, Cui2015, Cui2016, Wangj2016, Korasidis2017, LiWeiLing2017, Jain2018}, 
			Matching game \cite{Azougaghe2017, Ranjana2019},
			Three-sided stable matching \cite{Chu2017}, 
			Markov approximation \& stable matching \cite{CPham2017}, 
			College admission problem \cite{Kella2014, Xu2015},
			Resource Balancing Placement matchmaking \cite{Li2016} 
			Uncertainty principle of game theory \cite{Pillai2016}, 
			Egalitarian stable Matching \cite{Xu2011}, 
			Game theory \cite{ZhangY2017} \\ \hline
			MCDM & TOPSIS \cite{Tarighi2010, Tarighi2010a, Arianyan2016,Feng2019},  PROMETHEE \cite{Kabir2015}, General MCDM \cite{Rane2015},  Fuzzy  MCDM \cite{Araujo2018}, AHP  \cite{Son2019}   \\ \hline
			Heuristic & First-Fit \cite{Shi2011}, Next-Fit and First-Fit Decreasing \cite{Dow2016}, Opportunistic-Fit \cite{Georgiou2013},  Multiple Heuristics \cite{SunilRao2015}  \\ \hline
			Other & Fair proportional scheduler \cite{Chen2010} Bin packing \cite{Chen2017,He2011}, ILP repacking algorithm \cite{Sedaghat2013}, Distributed bartering algorithm \cite{Sonnek2010} Random search (biased statistical sampling) \cite{Tantawi2016} Random search (biased statistical sampling) \cite{Unuvar2014} Brute-force \cite{Viswanathan2011}, Constraints-based model \cite{Wright2012}, NA \cite{Zhou2013}, A framework \cite{Ooi2012, Menzel2012}, cloud broker \cite{Amato2013} \\ \hline
		\end{tabular}
\end{table*}

\subsection{Metaheuristic-based Approaches}
\subsubsection{Genetic \& Evolutionary Algorithms}
Evolutionary Algorithms (EAs) \cite{coello2007evolutionary} have proven to be a successful approach in solving many optimization problems including the VMP problem. EAs search for solutions by producing a population of candidates in every run. The population \emph{evolves} by applying crossover and mutation operators.  In the past decade, several preference-based EAs have been proposed (see for example \cite{Rachmawati2006} for a survey on preference-based EAs) where the goal is to incorporate users' preferences and desires when solving the problem. Thus, returning solutions that are not only optimized but also preferred by the user.

For VMP, preferences have been expressed as a utility function of the administrator for different combinations of SLA satisfaction, reliability and energy \cite{Deng2014} and as tie breaking mechanism for final solutions of the MOP \cite{Liu2017}. Also, preferences have been used to correlate two VMs to be placed on the same PM based on historical experience; thus, minimizing number of active PMs \cite{Liu2018}.	
	Indeed, one of the most well-suited roles of preferences in EA is to reduce the final Pareto set. Thus, the work of  \citet{Ihara2015}, \citet{Blagodurov2015} and \citet{Lopez-Pires2017} considered administrator preferences to reduce the final solutions of the VMP problem. A decentralized data center over different local departments have been considered in \citet{Saber2018} where it assumes every department administrator has a (possibly conflicting) preference.  \citet{Alashaikh2019} considered the case where some VMs have a ranking over which PM to be placed and used a variant of NSGA-II to return preferred solutions based on the notion of ceteris paribus preferences. 

\begin{table*}[htbp]
\caption{Detailed overview of the existing literature (Metaheuristics)}
\label{tab:details1}
\centering
\resizebox{\textwidth}{!}{\begin{tabular}{|m{2cm}|c|m{2cm}|m{3cm}|m{2cm}|m{2cm}|m{6cm}|}
\hline
{\bf Reference}&{\bf Perspective}&{\bf Problem Type}&{\bf Objectives} &{\bf Algorithm} &{\bf Preferences} &{\bf Comments}\\ \hline
 \citet{Addya2017}  &     cloud provider     &  placement &     Energy -- Revenue&  Simulated annealing  & Objectives &  A bi-objective optimization model to maximize revenue and minimize energy consumption. A trade-off between the two objectives is maintained based on DM's preference.  \\ \hline 
 \citet{Alashaikh2019}  &     cloud provider     &  placement&  Traffic -- Energy -- Resource wastage  &  MOEA  & Ordering &  Placement decision preferences are given a priori as a total order. \\ \hline 
 \citet{Blagodurov2015}  &     cloud provider     &  placement &  Traffic -- Energy -- Resource consumption &  NSGA-II  &     Objectives &  Posteriori preferences on objective space. \\ \hline 
 \citet{Deng2014}  &     cloud provider     &consolidation     &     Energy -- Resource consumption -- Availability/Reliability&  MOEA - improved grouping genetic algorithm  &     Objectives &  Defines a weighted utility functions as a weighted combination of SLA satisfaction, reliability and energy impacts, where the weights on each metric represents the DM preferences.  \\ \hline 
\citet{Ihara2015}  &     cloud provider     &  placement   migration&  Traffic   Energy -- Revenue -- No. of migrations -- Migration cost    &  Metaheuristics Memetic Algorithm  & Ordering &  Set of Pareto solutions with preferences applied on objective space. Multiple selection strategies such as distance to origin and Lexicographic Order (Each objective function is given in an order of evaluation) \\ \hline 
 \citet{Liu2017}  &     cloud provider     &consolidation     &     Energy -- No. of migrations &  Ant-colony  &  Inter-VM dependency  &  Obtains Pareto-based solutions and then the trade-off between the two objectives is adjusted based on the DM's preferences. \\ \hline 
 \citet{Liu2018}  &     cloud provider     &consolidation     &Resource consumption -- min. active PMs&  Ant-colony  &Inter-VM dependency&  The ant-colony algorithm  deposits the pheromone between VMs rather than between a VM and a PM and represents the preference between VM and PM. The preference is based on the historical experience of packing the VM together with those VMs that have already been placed on the PM. \\ \hline 
\citet{Lopez-Pires2017}  &     cloud provider     &  placement &  Traffic -- Energy -- load Balancing -- Revenue -- QoS priority&  Metaheuristics Memetic Algorithm  & Objectives &  Preference knowledge is utilized to reduce search space by imposing upper and lower bounds on objectives. \\ \hline 
{\citet{Mahalingam2015}}&  user -- multi-cloud  &  placement &satisfaction &  Simulated Annealing-based  & Sharing &  The model allows customers to specify their preferences (along with other QoS requirements) as hints based on whether their VMs can be located with other users? VMs or alone. Further, it considers soft constraints that only contribute to the satisfaction score and not enforced as the hard ones. (Uses Nefeli  \cite{Tsakalozos2013}) \\ \hline 
\citet{Saber2018}  &multi-cloud  & re-placement  &Energy -- Cost -- Migration cost -- Availability   &  Metaheuristics  &     Objectives &  Hybrid cloud managed by multiple entities, each of which has different requirements and preferences in maintaining its virtual cloud.  Preferences are incorporated as weights on the objectives. Three algorithms are utilized greedy algorithm to find initial solutions, NSGA to diversify, and a local search to obtain more non-dominated solutions.  \\ \hline
\citet{Tsakalozos2011}  &  user&migration& other/unspecified   &  Simulated annealing-based \& 2-phase heuristics  &    other/unspecified  &  Adopts hints to represents user preferences, internal physical cloud infrastructure, and high-level management goals that the DM seek. Fine-grained constraints refer to a) all specifics regarding the hosting capacity including features and resource availability in PMs, b) user-provided deployment hints and c) cloud administrative goals. A ranking algorithm finds different sets (Cohorts) of PMs that satisfy VM requests in terms of resources. Simulated Annealing algorithm to select a single cohort that satisfy preferences. \\ \hline
 \citet{Tsakalozos2013}  &     user -- cloud provider &  placement & other/unspecified   &  Simulated annealing  &    other/unspecified  &  proposes Nefeli, a hint-based VM placement algorithm.  In Nefeli, a predefined set of preferences are mapped onto a utility function with some weight (a score) on each preference criterion. The model also utilizes soft and hard constraints for both consumer and provider. \\ \hline 
\end{tabular}
}
\end{table*}

\subsubsection{Simulated Annealing}
	
	Another common technique to solve the VMP problem is Simulated Annealing, a single-based metaheuristic technique to solve optimization problems \cite{johnson1989optimization}. 
	A number of proposals have developed simulated annealing-based heuristics to solve the VMP problem. 
	\citet{Tsakalozos2011} developed a two-phase VM-to-PM placement scheme in which a group of PMs, termed cohort, are first identified as potential host based on some properties, then the final constrained mapping is resolved taking into account customers' special requirements and preferences. The goal of selecting a subset of all available PMs in the first phase is to reduce the number of constraints and the search space during the second phase. 
	In addition to resource and cloud specifications constraints, the scheme considers customers' deployment hints such as colocation of multiple VMs in a single PM and reserving a whole PM for a single VM. The authors proposed two heuristics for filtering cohort PMs and placement based on simulated annealing. 

	Moreover, \citet{Tsakalozos2013} developed a cloud gateway, Nefeli, that takes customers' deployment hints when mapping incoming VMs to PMs. 
	The hints realize several preferences related to VMs deployment such as colocation, availability, proximity and other information. 
	
	Their objective is to optimize cloud energy efficiency and load balancing while meeting cloud admin constraints. 
	At the same time, Nefeli tries to satisfy customers' preferences but it may only satisfy parts of them and ignore the rest based on the available resources. 
	For evaluating multiple placement solutions, Nefeli adopts a weighted score function on constraints, where a weight derived from customers' hints is associated with each constraint. 
	The set of constraints is further decoupled into hard and soft constraints. While the first is strict and has to be met, the second only contributes to the score function. 
	Eventually, the optimal solution is defined as the one with the highest score. 
	The algorithm adopts a simulated annealing-based algorithm to solve the placement problem and a search space reduction method was used for large scale problems. 

	\citet{Mahalingam2015} adopted Nefeli to develop a cloud service selection mechanism that takes into account VMs' preferences for price and QoS and search for acceptable offers. 
	Lastly, \citet{Addya2017} developed a bi-objective optimization model to maximize revenue and minimize energy consumption and solve it using a simulated annealing technique. The proposed model attempts to maximize the revenue and minimize the power budget. A trade-off between the two objectives is maintained based on administrator's preference. 

\Cref{tab:details1} provides a summary to the works presented here. 

\subsection{Matching}
VMP can be naturally described as a matching problem.  The set of participants in this matching are often the available VMs and PMs where every participant expresses their preferences over who to partner with. Such preference relations are due to some metrics such as transmission cost or processing capabilities and requirements which can be measured or estimated. The result of the matching is that every VM is placed on exactly one PM. Stability is an important property where a matching is considered \emph{stable} when no machine (VM or PM) would prefer another machine to its current partner \cite{Phuke2018}. 

The literature shows different roles of preferences when matching VMs to PMs. VMs could rank PMs based on some properties such as transmission cost \cite{Xu2011}, migration time and cost  \cite{Kella2014} , processing capabilities  \cite{Xu2015} , and security  \cite{Azougaghe2017} while PMs rank VMs based on their migration overhead \cite{Kella2014} , CPU and resource utilization \cite{Xu2015, Chu2017}, and risk  \cite{CPham2017, Azougaghe2017}. \citet{Kim2014} also considered the case where preferences are not necessarily a ranking and ties are allowed. The goal is often to minimize the overall dissatisfaction score which is defined as sum of ranks. This score can be intuitively seen as the average happiness of VMs and PMs.

	Preferences have been utilized to place high throughput applications near computation nodes to minimize the cost of communication  \cite{Korupolu2009}.  \citet{Dhillon2013} addressed consolidating multiple VMs on the same PM while minimizing degradation performance.  
	 It assumed that every VM acts as a selfish agent and has a preference over which VMs to partner with on the same PM. The preference is based on some hidden cost function related to the VM.  \citet{Kella2014} considered the case of multiple cloud providers with intermediate brokers where customers express preferences. The live migration problem has been formulated as a matching problem with the goal of reducing the overall energy consumption without degrading QoS.  \citet{Cui2015} considered networks where the policies, which reside on middle boxes such as firewalls and load balancers, etc., are migrated first then a matching between PMs and VMs based on the communication cost and resource usage is applied. In \cite{Li2016}, a descriptive language based on XML for the preferences of data center managers was proposed. The goal is to automate the matching procedure while classifying preferences into two classes; namely, flexible and inflexible. 
\citet{Wangj2016} addressed the problem when VMs and PMs cooperate on a mutual objective (reducing the overall energy consumption).  \citet{Chu2017} added storage PMs as a third set into the matching where storage PMs prefer smaller VMs.  \citet{Pillai2016}  considered the case where IaaS request is beyond the capability of a single PM and therefore developed a mechanism to form coalition with the goal of minimizing the total execution time based on uncertainty principle in game theory. The work in \cite{ZhangY2017} addressed the problem of migrating VMs for security preferences to make it harder for attackers to locate targeted VMs. Lastly, \citet{Korasidis2017} enhanced the scalability of the stable matching technique for the VMP problem by considering only the top-k alternatives in the preference list. \Cref{tab:details2} summarizes the presented works of this method. 

\begin{table*}[htbp]
\caption{Detailed overview of the existing literature (Matching)}
\label{tab:details2}
\centering
\resizebox{\textwidth}{!}{\begin{tabular}{|m{2cm}|c|m{2cm}|m{3cm}|m{2cm}|m{2cm}|m{6cm}|}
\hline
{\bf Reference}&{\bf Perspective}&{\bf Problem Type}&{\bf Objectives} &{\bf Algorithm} &{\bf Preferences} &{\bf Comments}\\ \hline
 \citet{Azougaghe2017}  &     cloud provider     &     migration&     Energy -- Resource consumption &  Matching game  &  Resource -- Cost -- Security     &  VMs preferences and PMs are represented as utility functions that take into considerations multiple attributes such as resources and security level for VMs and risk, energy, and revenue for PMs. \\ \hline 
 \citet{Chu2017}  &     cloud provider     &  placement &Cost   &  Three-sided Stable matching  &  Resource -- Distance    &  PM nodes are decoupled as computing and storage nodes. three-sided matching among computing, storage resources and applications, each with its preference list \\ \hline 
 \citet{CPham2017}  &     cloud provider     &  placement &  Traffic   Energy    &  Markov approx. and Stable matching  &  Resource -- Inter-VM dependency&  vNFs of the same service chain maintain a preference list on PMs, while each PM maintains a preference list on vNF instances. \\ \hline 
 \citet{Cui2015}  &     cloud provider     &     migration   consolidation     &Resource consumption -- Policies     &  Stable matching  &  Resource    &  preference knowledge is the cost of paths and links for VMs traffic. \\ \hline  
 \citet{Cui2016}  &     cloud provider     &     migration   consolidation     &Resource consumption -- Policies     &  Stable matching  &  Resource    &  same as above. \\ \hline 
 \citet{Dhillon2013}  &     cloud provider     &  placement &Satisfaction &  Stable matching  & Colocation     &  Every VM acts as selfish agent and has a preference (strict total order) over other VMs to partner with. The system tries to minimize the degradation of VMs placement problem and find a solution that minimizes the global performance degradation. \\ \hline 
 \citet{Ranjana2019}  &cloud provider  &  placement &Energy --comm. cost -- min. active PMs &  Stable matching  &   VM dependency   &  Affinity-based algorithm. \\ \hline  
 \citet{Jain2018}  &multi-cloud  &  placement & &  Stable matching  &   Cost   &  multiple cloud providers, intermediate brokers and end users \\ \hline  
 \citet{Kella2014}  &multi-cloud  &     migration&     Energy    &  Stable matching /college admission problem  &  Resource -- Migration cost&  VMs preference list contain PMs put in an ascending order according to the transfer time. PMs preference list contain VMs put in an ascending order according to their impact to cpu utilization. \\ \hline 
 \citet{Kim2014}  &     cloud provider     &     migration& &  Stable matching  &  Distance    &  the model allows ties in the preference list ordering (\ie preferences are not given in a strict order) \\ \hline 
 \citet{Korasidis2017}  &  user -- cloud provider     &  placement & other/unspecified   &  Stable matching  & other/unspecified  &  Addresses algorithm scalability. \\ \hline 
 \citet{Korupolu2009}&cloud provider&placement &Cost   &  Stable matching&Resource& One of the first proposals that applied stable matching to the VM placement problem. It considers compute and storage PMs, and applications of different computation and data needs. Requirements and other applications preferences can be integrated into a weighted utility function.  \\ \hline 
 \citet{Li2016}  & user &consolidation   & min. active PMs -- Resource wastage  &  Stable matching (resource balancing and matchmaking)  & Location -- Specific PM&  Defined normal and categorized types of preferences and flexible and inflexible preferences. \\ \hline 
 \citet{LiWeiLing2017}  &     cloud provider     &consolidation     &     Energy    &  Stable matching  &Migration cost -- Resource&  Modeled as a matching problem on a weighted bipartite graph, with VM migration cost used as a weight on edge between VM and PM. Each VM candidate for migration has a preference list on a subset of PMs chosen to accommodate migrated VMs. The list is ordered based on weights.  \\ \hline 
 \citet{Pillai2016}  &     cloud provider     &  placement & Revenue&  Stable matching and uncertainty principle of game theory  &     Location &  The proposal considers forming coalition between PMs to satisfy requirements that are beyond single machine capabilities. The PM has preference over which coalition of machines to go with based on expected payoff. \\ \hline 
 \citet{Wangj2016}  &     cloud provider     &     migration&     Energy    &  stable matching  &  Resource    &  PMs and VMs have mutual objective of freeing overloaded PMs and utilizing underutilized PMs in order to reduce overall energy consumption. The proposed algorithm tries to achieve a reasonable trade-off between power consumption and SLA violation. Preference list is computed accordingly.  \\ \hline
 \citet{Xu2011}  &     cloud provider     &     migration& other/unspecified   &  Egalitarian stable Matching  &    other/unspecified  &  A VM or a server ranks the agents on the other side based on some criteria, which is its preference list. An algorithm then computes a matching that cannot be improved by any pair of VM and server. VMs rank servers in an ascending order of the transmission cost. Servers rank VMs in an ascending order of the migration overhead. \\ \hline 
 \citet{Xu2013}  &  user   cloud provider     &  placement &Resource consumption -- other/unspecified   &  Stable matching  &    other/unspecified  &  A VM preference list ranks servers according to their resources. PMs may prefer packing larger or smaller VMs recognizing DM's preference of consolidation or load balancing, respectively. Proposed a revised DA algorithm that finds weakly stable matching which can be an effective approach for handling large-scale problems. \\ \hline 
 \citet{Xu2015}  &  user   cloud provider     &  placement &Resource consumption &  Stable matching/ College admission problem  &  Resource -- Inter-VM dependency  &  The stable matching theory is applied to generate an optimal mapping from containers to physical servers. \\ \hline 
 \citet{ZhangY2017}  &     cloud provider     &     migration&Cost   &  game theory  & Security     &  Users security preferences are modeled as a linear function of the system diversity (using weighted function of the total number possible VM-placements). The cost accounts for migration time and number of migrations. \\ \hline 
\end{tabular}
}
\end{table*}

\subsection{MCDM}
\label{sec:mcdm}
Multi Criteria Decision Making (MCDM) \cite{keeney1993decisions} is a field of study to find and analyze preferred options in complex problems under conflicting criteria. Two well-known techniques in MCDM are AHP and TOPSIS \cite{Whaiduzzaman2014}.  \citet{Araujo2018} presented an MCDM method to rank the best cloud infrastructure based on a customized dependability metric and the cost. The customized metric takes customer service constraints such as reliability, downtime, and cost into consideration. The model allows users to express their preferences on cost versus availability, through a planning and analysis tool, to obtain the best configuration for them. \citet{Tarighi2010,Tarighi2010a} presented a method for VM migration between cluster nodes (a group of PMs) using TOPSIS algorithm
The aim of their algorithm is to achieve load balancing among PMs by moving load from the most overloaded PMs to the least-overloaded PMs and attempting to minimize migration cost defined in terms of amount of data copied and transferred. 
The TOPSIS method is a multi criteria decision making technique. It stands for order preference by similarity to ideal solution. 
It is based on selecting an alternative solution, from a finite set of alternatives, 
that is closest to the ideal solution and farthest from the negative ideal solution. 
The ideal solution is the solution that maximizes the benefits and minimizes the cost, whereas the negative ideal solution does the opposite.  
Typically, each alternative is evaluated with respect to some attributes (decision criteria), and each attribute is assigned a weight based on its importance (\ie given by the user). 

The alternatives are ranked according to their relative closeness to the ideal solution. The closer the alternative, the better. 

PM nodes are sorted from the highest overloaded to the idle ones based on some criteria such as CPU, RAM, NET, etc. Each attribute is given a weight. Then after finding the most saturated PM, the algorithm tries to find the best VM candidate for migration based on similar criteria w.r.t. VMs. 
The proposed TOPSIS algorithm can receive three types of information including deterministic, linguistic, and fuzzy information. 

\citet{Arianyan2016} proposed a TOPSIS algorithm aiming to optimize energy, SLA, and number of migrations in a cloud data center. 
The algorithm selects VMs to be migrated when overloaded PMs are predicted. 
It considers a decision matrix of the resource utilizations and capacities of VMs. 
Each VM is assigned a score for each resource criterion based on the ideal points concept. 
The total score of a VM is the weighted combination of the scores of each criterion. The weights for each VM are set independently. Similarly,  \citet{Feng2019}  developed a TOPSIS algorithm that incorporates AHP to find weights of coefficients of the resources decision matrix. 
 \citet{Rane2015} introduced a cloud broker architecture to help application customers select their cloud provider considering both functional and non-functional requirements on the service. 
While functional requirements must be satisfied, non-functional requirements are associated with weights to specify their priority on these requirements. 
Further, an MCDM constraint significance factor realizing the preference of each constraint is utilized. 
Then, the method tries to maximize customers' satisfaction.
\citet{Kabir2015} proposed a two phase VMP algorithm that first selects a cluster of PMs and then a PM inside the cluster to place a VM. The proposed algorithm is based on the MCDM method PROMETHEE. 
Overall, the method is based on outranking alternatives. 
Preferences on each criterion are quantified as real numbers and aggregated onto a weighted preference index. 
Then, each alternative is evaluated by the number of alternatives that outrank it and the number of alternatives it outranks based on the preference index. The alternative with the largest net outranking is the most preferred one. 

\Cref{tab:details3} summarizes the works presented here. 

\begin{table*}[htbp]
\caption{Detailed overview of the existing literature (MCDM)}
\label{tab:details3}
\centering
\resizebox{\textwidth}{!}{\begin{tabular}{|m{2cm}|c|m{2cm}|m{3cm}|m{2cm}|m{2cm}|m{6cm}|}
\hline
{\bf Reference}&{\bf Perspective}&{\bf Problem Type}&{\bf Objectives} &{\bf Algorithm} &{\bf Preferences} &{\bf Comments}\\ \hline
\citet{Araujo2018} & user & placement& Availability -- Cost&Stochastic model \& MCDM & Objectives & The authors use stochastic models to assess and evaluate different clouds from user's perspective. Assessment results are then used to rank solutions, taking into account the user's preference of each criterion (set as a weight). \\ \hline
 \citet{Arianyan2016}  &     cloud provider     &consolidation     & other/unspecified  &  TOPSIS  &    other/unspecified  &  Preferences are expressed as weights for each criteria. The weights are computed based the average utilization of all
system resources in the data center (\ie functional). The utmost objective is to remove hotspots. \\ \hline 
 \citet{Feng2019}  &  cloud provider& consolidation  &Cost &  TOPSIS/AHP  &    Resources  &  The algorithm is mainly concerned about the migration decision.  \\ \hline
 \citet{Kabir2015}  &multi-cloud  &  placement &Resource consumption &  PROMETHEE  &    other/unspecified  &  a preference function is used to denote the preference of solution over another for each criterion. Then, each solution is evaluated by the number of alternatives outrank it and the number of alternatives it outranks based on the preference index. The solution with the highest net outranking flow is considered as the most preferred.  \\ \hline 
 \citet{Rane2015}  &  user& re-placement  &Satisfaction &  MCDM  &    other/unspecified  &  Defines a constraint significance factor that signifies the preference of customer corresponding to each constraint.  \\ \hline
 \citet{Son2019} &   user -- cloud provider     &   placement/ migration& Resource utilization -- QoS -- no. of migrations -- Energy&  MCDM/ AHP  &Objectives & Proposes to use Analytical Hierarchy process (AHP) coupled with a fuzzy interferer system to solve the VMP problem. By monitoring resources and analyzing service requirements, a placement decision can be obtained from the fuzzy system. Then through AHP algorithm, a placement action is performed. The AHP is managed to provide analysis of the weights for each metrics involved in the decision.  \\ \hline 
 \citet{Tarighi2010}  &cloud provider&     migration&    load Balancing Migration cost    &  Fuzzy TOPSIS  &  Resource    &  deterministic, linguistic, and fuzzy preference information. \\ \hline 
 \citet{Tarighi2010a}  &cloud provider&     migration&    load Balancing Migration cost    &  TOPSIS  &  Resource    &  same as other one \\ \hline 
\end{tabular}
}
\end{table*}

\subsection{Heuristics}
\label{sec:heuristics}
%
\citet{Shi2011} proposed a First-Fit algorithm aiming to maximize the profit without violating the SLA and the power budget. 
The algorithm considers two kinds of SLAs: the hard SLA and the soft SLA. 
A hard SLA is not tolerable and violating any of its terms yields no revenue. 
For a soft SLA, application customers can tolerate performance degradation to some extent and provider revenue suffers certain loss accordingly. 
The more elastic the VMs are, the more VMs can be consolidated onto a PM.
Thus, this elastic model allows providers to manage revenue based on SLA.
Moreover, \citet{Dow2016} used modified Next-Fit and First-Fit heuristics to solve the VMP pin packing problem. They proposed to decompose the VM consolidation into two phases; namely, an abstract packing phase that satisfies constraints, and a concrete assignment phase based on preferential colocation and anti-colocation. 
Alternatively, \citet{Georgiou2013} developed an Opportunistic-Fit algorithm addressing the problem of placing VMs with an Infrasturctue-as-a-service (IaaS) request, in which a group of VMs are jointly consolidated in the cloud. 
Considering DC PortLand architecture, they proposed two VM placement algorithms that exploit users' hints about the desired features of the requested IaaS.
Users specify their requirements on the VMs and their colocation preferences through an XML document. 
The hints considered here are concerned with the bandwidth needs for specific pairs of VMs and the anti-colocation constraints for pairs of VMs. 
In the first algorithm, a greedy algorithm places requested VMs as close as possible to each other while satisfying colocation requirements as much as possible; otherwise, a path with sufficient bandwidth is established between them. 
The second algorithm tries to minimize the time expended for the placement decision by identifying promising neighborhoods of PMs for deploying the IaaS first before applying the greedy placement algorithm. 

\subsection{Other Methods}
\label{sec:othermethods}
In this part, we review several works that adopt exact methods or general customized heuristics. 

\citet{Viswanathan2011} addressed the placement of High-Performance Computing (HPC) applications in clouds. 
They proposed a VM consolidation strategy that aims at optimizing resource utilization and energy efficiency while satisfying application's QoS. 
A brute-force search algorithm is used to optimize energy efficiency and performance with a parameter utilized to adjust the possible trade-off between them. 
\citet{He2011} proposed a matching algorithm for VMs migration addressing IaaS management in clouds.  
For incoming IaaS request, the algorithm finds a suitable PM based on available resources. 
If there are multiple suitable PMs, the system gives preferential attention to more important customers. 
A PM is recognized by the different properties it has, and a customer may express different preferences over these properties. The preferences are given in values indicating their importance to the customer. 
Then, the algorithm chooses the most suitable PM that satisfies the preference order the most. 
\citet{Wright2012} developed a two phase resource selection model that can handle a combination of hard and soft constraints. The model intends to enable users to match their application requirements to infrastructure resources. A direct search algorithm is proposed to solve the problem.

\citet{Chen2010} considered the problem of mapping virtual CPUs of VMs to physical ones. 
A proportional scheduler (PS) is proposed to provide fair allocation of CPU resources among virtual CPUs. It assigns a weight to every VM based on its computational requirements. 
CPU-affinity is defined as a property of a virtual CPU that describes which physical CPU it can run on. 
In their study, they illustrated that the inappropriate assignment of weights may produce unfair proportional sharing of CPU. 
\citet{Sonnek2010} presented a decentralized affinity-aware migration technique that dynamically migrates VMs based on network affinity between pairs of VMs in order to minimize intra-cloud traffic. Affinity here is defined as VMs interdependency (\ie volume of traffic monitored between pairs of VMs). The authors developed a distributed bartering algorithm that strives to place VMs on a PM that is closer to the data they need. VMs have preference ordering on which PM they migrate to based on the monitored intra-traffic. The VM is responsible for negotiating the placement with PMs which improves the scalability of the algorithm and minimizes the overhead. 
In a similar vein, \citet{Ranjana2019} classify PMs based on CPU and communication bandwidth and then apply affinity-based matching to minimize energy consumption, number of active PMs, and communication cost. 
\citet{Zhou2013} developed three live VM storage migration mechanisms. 
The mechanisms introduce a trade-off between the efficiency and cost of VM migration. 
They also consider user experience (IO penalty), cluster management (migration time) and device usage (degree of wear) in the migration decision by utilizing a migration cost (utility function) in which weights reflect different design preferences.

 \citet{Tantawi2016} proposed a biased random search method based on biased statistical sampling. It maps VMs and other workloads into PMs and other physical entities so that an objective function is optimized. 
The algorithm takes into account both administrators' and users' requirements and allows users to specify their preferences over the constraints (hard/soft constraints) over some criteria such as location. 
The biasing pushes the solution towards satisfying constraints and preferences. 
\citet{Unuvar2014} adopted the same method to address VMP in hybrid clouds. The placement decision takes into account whether private or public cloud is preferred for a customer and a customer preference on splitting VMs over private and public clouds. 
\citet{Sotiriadis2016} presented VMs migration within OpenStack cloud platforms considering the FIWARE platform \footnote{FIWARE is developed within the FI-STAR FP7 project that utilizes cloud technologies and IoT to support healthcare services and applications, http://www.fi-star.eu.}. The migration tool is developed following the FI-WARE conceptual model and offers a set of software modules, called Specific Enablers (SEs), that allow flexible development of healthcare applications. 
Within the deployment process of VMs, the system stores and retrieves users' preferences including geographic location preference. 
This is done utilizing a graphical interface to configure customer preferences.
\citet{Chen2017} proposed a Joint Affinity Aware Grouping and Bin Packing (JAGBP) method that aims to minimize application performance cost and maximize resource utilization. The affinity relationships of VMs define colocation and disperse placement preference based on VMs interdependency. 
They define three affinity types and affinity relations; namely, colocation placement affinity/relation (co-affinity/relation), no-colocation placement affinity/relation (no co-affinity/relation) and fixed placement affinity/relation (fixed-affinity/relation).
JAGBP was empirically found to improve the above objectives compared to non-affinity aware placement. 

\Cref{tab:details4} summarizes the works presented in \cref{sec:heuristics,sec:othermethods}. 

\begin{table*}[htbp]
\caption{Detailed overview of existing literature (Heuristics \& others)}
\label{tab:details4}
\centering
\resizebox{\textwidth}{!}{\begin{tabular}{|m{2cm}|c|m{2cm}|m{3cm}|m{2cm}|m{2cm}|m{6cm}|}
\hline
{\bf Reference}&{\bf Perspective}&{\bf Problem Type}&{\bf Objectives} &{\bf Algorithm} &{\bf Preferences} &{\bf Comments}\\ \hline
\citet{Chen2010}  &     cloud provider     &  placement &Resource consumption &  Heuristics  &  Resource    &  VMs are classified into domains based on their CPU-affinity. \\ \hline 
  \citet{Chen2017}  &cloud provider &  placement & min. active PMs&  Bin packing  & Colocation -- Anti-colocation  &  Affinity-aware placement -- hard constraint \\ \hline 
\citet{Dow2016}  &     cloud provider     &consolidation     & min. active PMs&  Heuristics (pin-packing using next-fit and ffd)  & Colocation -- Anti-colocation  &  decomposed into two phases; resources computation then colocation. \\ \hline
\citet{Georgiou2013}  &     cloud provider     &  placement & other/unspecified   &  Heuristics (Opportunistic-fit greedy algorithm)  &  Resource -- Anti-colocation  &  IaaS requests. Used the notion of hints to allow users specify their bandwidth needs for specific pairs of VMs and  anti-collocation constraints for pairs of VMs. aim to faster solution time. Flexibility of constraints. \\ \hline 
\citet{He2011}  &  user --  cloud provider     &     migration&Resource consumption -- Cost   &  Heuristics  & Ordering &  Each PM machine has a set of properties associated with its resources and attributes (\eg RAM, type of OS, etc). If multiple PMs are suitable for hosting a VM (\ie satisfy the requirements), then the system looks for user's preferences. \\ \hline 
\citet{Ooi2012} &cloud provider&placement -- replication&Availability&Team formation&other/unspecified& A framework for resource evaluation to identify preferable resources and prepare for service placement.  \\ \hline
\citet{Sedaghat2013}&cloud provider & repacking& Resource utilization& ILP & other/unspecified & Repacking by changing the configuration of VMs (either number or sizes) that reside in a server, as the load changes. The goal is minimizes resource utilization cost accounting for both price and performance issues. A decision of repacking or not can be based on the calculated cost and other factors \eg expected time to next change, interruption time, etc. \\ \hline
\citet{Shi2011}  &     cloud provider     &  placement &     Energy -- Revenue&  First-fit Heuristic  &SLA&  Hard SLA and soft SLA. \\ \hline 
\citet{Sonnek2010}  &     cloud provider     &     migration&  Traffic &  Heuristics  &Inter-VM dependency -- Colocation     &  affinity; satisfy it as much as possible \\ \hline 
\citet{Sotiriadis2016}  &  user -- multi-cloud  &     migration& other/unspecified   &  Migration tool  &     location -- other/unspecified  &   migration of VMs in FIWARE systems \\ \hline 
\citet{SunilRao2015}  &&     migration& min. active PMs -- Resource wastage  &  Heuristics  & Sharing &not implemented/discussion only.\\ \hline 
\citet{Tantawi2016}  &  user -- cloud provider     &  placement &    Load Balancing     & Heuristics/random search (biased statistical sampling)  &     Location -- Colocation     &  location constraints can be hard or soft. \\ \hline 
\citet{Unuvar2014}  &  user -- multi-cloud  &  placement &  Traffic  load Balancing  Cost   &  Heuristics - Biased Importance Sampling  &  Resource -- Cost -- other/unspecified  & Introduced a hybrid cloud placement algorithm that incorporates two policy parameter: reservation and cost weight and addresses the trade-off between deployment cost and request rejection rate.   \\ \hline 
\citet{Viswanathan2011}  &     cloud provider     &  placement &     Energy -- Completion time&  Brute-force + set partitioning algorithm  & Objectives &  Addresses HPC applications placement in DC. \\ \hline 
\citet{Wright2012}  &  user&  placement & other/unspecified   &  Heuristics  &    other/unspecified  &  user perspective. measured/evaluated in cost \\ \hline 
\citet{Zhou2013}  & cloud provider &  migration&Cost -- Migration cost    &  Heuristics  &    other/unspecified  &  Incorporates user experience (IO penalty), cluster management (migration time) and device usage (degree of wear). Utilizes a weighted cost function in which the weights on each metric reflects different design preferences. \\ \hline
\end{tabular}
}
\end{table*}

\subsection{Discussion}
It is clear from the five classes of approaches mentioned above  that there exists different ways to incorporate preferences within the problem with different goals as some methods try to reduce the search space (e.g., bounds on objectives \cite{ihara2015many}, preference order on the assignment \cite{Son2019}) while others generate a single solution or an ordered list of solutions (e.g., most of matching algorithms). Metaheuristics are viable option in tackling complex domains with conflicting objectives. For instance, when reducing the energy while maximizing the revenue during virtualization. However, they are prone to returning suboptimal solutions and require good tradeoff between exploration and exploitation that, in practice, is hard to construct efficiently. In case the preferences are user-centric, \ie they represent provider or user, there is a need for interactive metaheuristics approaches  that consult the user during the search process \cite{Lopez-Pires2017}.

Indeed, the heuristic approaches require domain knowledge to construct an effective heuristic function with which promising search space is discovered first. Such domain knowledge is sometimes prohibitive when multiple stakeholders' preferences are involved. One needs to assume preferences follow certain structure and often construct a heuristic function for a simplified version of the original problem. Hence, it is obvious that the proposed heuristic approaches are general and usually domain independent \cite{Addya2017}.

Most of the matching methods consider resource as the attribute to base their strategy on. This is a reasonable consideration as physical machines have limited resources and utilizing them efficiently when looking for a stable matching between VMs and PMs is of great importance. Stability of the match guarantee the average happiness among all participants, \ie that there exists no pair $(X,Y)$ of virtual machine and physical machine that prefer to be with each other than their current partner (\eg  the physical machine hosting $X$, and the VMs placed on $Y$). 

MCDM approaches such as AHP and TOPSIS are known to require large number of pairwise comparisons and do not show attribute interdependencies. They are also time consuming and have scalability issues when the number of attributes is large. In such case, preferential dependencies among attributes can lower the time needed to assess preference between alternatives \cite{KeeneyRaiffa76}. 

\section{Benefits, Research Challenges, and Trends}
\label{sec:benefits}
\subsection{Potential Benefits}
After the detailed review of preference applications, here we draw insights regarding the benefits of incorporating preferences into the VMP problem.

\begin{description}
\item[User Satisfaction]
Indeed one of the main benefits of considering preferences is to find solutions that are not only optimized but also preferred by the user or decision maker. 
Apart from setting firm and solid requirements, a customer or a cloud administrator can express additional soft requirements through preferences. Considering such preferences would increase the satisfaction of both parties and usually a graphical interface is utilized to specify the preferences \cite{Sotiriadis2016,Tsakalozos2013,Tsakalozos2011}.

\item[Reducing the Complexity] 
VMP is a complex problem given all the possible conflicting objectives, constraints from multiple entities, functional and non-functional properties, random nature of workload changes, and managerial and geographical policies. 
It is remarkably noticed in the literature that considering preferences %
not only handles users' and provider's fine details but also %
introduces some sort of flexibility to the problem. 
In addition, the problem of optimizing VMP is known to be NP-complete and even sub-optimal solutions needs to be obtained in a timely manner. Preferences can effectively not only provide but also be the tool to control the margin of flexibility considering performance, efficiency, solution time \cite{roytman2013algorithm}, and violation rate \cite{Chowdhury2015}. 

For example, \citet{Ye2015} reported in their study that the number of PMs required to host VMs with colocation constraints decreases when the constraints are softened. 
In addition, the gain of incorporating preferences improves the scalability to large-scale problems as in \cite{Korasidis2017}. 

Moreover, preference information can be utilized to set criteria for managing complex processes. 
As an example, consider the dynamic PM consolidation process. 
Typically, there exists a trade-off between the goodness of the process and completion time due the complexity of the process \cite{MohamadiBahramAbadi2018}. Preferences can be utilized to resolve this trade-off in an efficient manner. 
In addition, cloud providers typically offer predefined instances of virtual resources with different capabilities as well as value-added features and services (\eg choice of OS, traffic on a private VLAN, etc). The customer choices impact the pricing of such leasing. \citet{Wright2012} defined application soft and hard constraints to reduce the complexity of choosing suitable deployment.

Besides, there are cases that inspire to use predefined soft constraints. 
\citet{Hao2017} considered bounded constraint violation to maintain and fix the feasibility of the LP problem under study. The selection of constraints and the extent to which they are softened is critical to the quality of the solution and should be set based on administrator's preferences rather than arbitrarily.  

\item[Reducing Search Space]

Multiobjective problems typically obtain multiple Pareto solutions, and the number of these solutions grows significantly as the number of objectives grows. 
Thus, preferences have been utilized to help mitigate this burden \cite{Deng2014, Ihara2015, Liu2018, Saber2018, Alashaikh2019,Lopez-Pires2017}. 
The same can be said about other algorithms (\eg alternative based algorithms). 
For example, the work in \cite{Tsakalozos2011} considered a two phase approach and applied non-functional preferences in the second phase to reduce the number of candidate PMs. 

 Another example, there can be the case where two or more  solutions are available (\eg ties in preference list \cite{Phuke2018}), and favoring such an option might have some advantages that the DM was not aware of.

\item[Adaptation of Specialized Services]

Preferences and soft constraints can be utilized within VMP schemes to enable the hosting and QoS guarantees of specialized services 
(\eg mapping special services to a PM or a set of PMs). 
Security concerns \cite{ZhangY2017}, NFV chaining \cite{CPham2017}, HPC placement \cite{Viswanathan2011}, mobile applications \cite{Zhang2016}, smart TV services \cite{Kim2014}, and Web services \cite{Menzel2012} are examples of VMP for specialized services that incorporate the notion of preferences in their approaches. 

\item[Improving Performance ] 
Preferences can be used as a guideline for optimizing the performance of a system. Such preferences are usually inspired from meta-analysis of the problem and identified as preferred settings or configurations to the machine. 
The work in \cite{Verma:2010} for instance, provides a detailed experimental study and identifies the parameters that influence the cost of migrating a VM and provides some recommended rules to improve VM migration efficiency. We expect that these rules (\eg migrate applications with small active memory first, migrate low priority applications first) can suitably be identified as preferences rather than strict constraints or objectives. 

Furthermore, constraints and objectives are considered known, hard and strict. 
preferences are soft, and can be given on top of the problem in a way that does not impact constraints and objectives. 
However, preferences can be known in advance, known a posteriori, or learned throughout the system's life. 
The later can be useful in the VMP context as the system preferences can be learned and adapted to the operational status of the cloud \cite{Battiti2010}. 
This gives a wider perspective to administrators on getting to know how different unknown options arise from the problem and to what extent they can change the output \cite{Battiti2010,Ooi2012}

\end{description}

 \subsection{Trends in the cloud}
 
 Software-defined Networking (SDN) and Network Function Virtualization (NFV) are two of the key technologies in the cloud and 5G networks. 
The SDN architecture provides flexible and programmable network control by decoupling the control plane from the data plane. 
This enables the use of simple network devices (\ie switches) to create and manage networks for different purposes on the fly via a logically centralized SDN controller.

NFV enables deploying network functions (\eg routing, firewalling, load balancing, etc.) as software applications (\ie VM) stored and run on generic servers (\ie PM). 
It also introduces flexibility in terms of location, in which functions can be placed anywhere in the network satisfying users and provider requirements. 
However for a specific task that requires more than function, there is an order sequence that must be followed \cite{Laghrissi2019}.

The advent of SDN/NFV technologies has facilitated the \emph{cloudification} of many communication systems \cite{Han2015,Mijumbi2016,Dai2017,Laghrissi2019}. 
Here we shed light on a number of trends that leverage virtualization and cloud computing technologies in order to improve overall system performance. 

\begin{description}

\item[Internet of Spectrum Devices] aims to expand the radio spectrum of wireless mobile networks in which devices/elements of the network (\eg base stations) have no restriction on spectrum use and can utilize multiple shared resources of the spectrum dynamically. 
A cloud is responsible for gathering spectrum status/info. from monitoring devices and distributing them to spectrum utilizing devices. 
There is a virtual DM for each entity in the spectrum cloud that is responsible for making spectrum decisions. 
Various cloud computing operations can be applied based on DM request and users' preferences
\cite{WuQ2016}. 

\item[Cloud-Radio Access Networks (C-RANs)] offer cloud-based architecture in which all control algorithms are placed and executed by data centers \cite{Wu2015}. 

Specifically, in a C-RAN, traditional base stations functionalities are decoupled into radio functions that are stored in the remote radio unit (RRU) and digital and higher layer functions that are stored in the baseband unit (BBU). The RRU remains at the base station site, while the BBU is moved to the cloud. 
C-RANs extend the mobile radio access capacity and coverage by pooling BBU resources from multiple cells. 
These cells can be heterogenous and may belong to multiple collaborating and competing carriers. 
Following NFV standards, each BBU functionality is represented by one or more VMs which can be enabled or disabled for each base station. 
Further, these VMs have to be placed in PMs that satisfy their requirements
and should be dynamically migrated to where they are needed.
The sequence and order of execution of network functions as well as the colocation requirements have to be taken into account when making a placement decision. 

\item[Content delivery and Caching]
 
In modern ISP networks, content (videos and cached pages) can be stored in VMs, and the CDN is configured adaptively by SDN controllers to account for traffic loads on the ISP network \cite{Wichtlhuber2015}.

\item[Small Cells] 

One of the challenges to small cell technologies (\eg WiFi and Femtocells) is high-level of interference between cells due to the lack of channel coordination. Cloud-based control is a promising method to manage channel assignment between different, multi-vendor, small cells. 
%
There are many other technologies that rely on cloud computing concepts such as mobile cloud, edge Computing, cognitive IoT, etc. 
In general, cloud-based architectures can provide more flexibility and centralized control and management of the participating entities/devices, in which, coordination and performance can be optimized. 
Furthermore, proper understanding and translation of the requirements of each VM (\eg application) will lead to desired placement decisions.

\item[VMP with Preferences on the Edge]
The problem of solving VMP with preferences on the edge brings several challenges on top of the VMP challenges. One needs compact models to represent and reason about preferences. Such models need to be compact and succinct in their preference language to handle the possibly large number of attributes and limited resources on the edge. Even in the case of successfully stored, preferences on the edge may be insecure for the genuine user's preferences and selection criteria and, therefore, privacy-preserving techniques for storing and dealing with preferences on the edge are required. Recently, \citet{Ouyang2019} proposed an adaptive algorithm for service placement on the edge, which jointly optimizes latency and migration cost, weighted by the on-line learned user's preferences. \citet{Qian2019} developed an privacy-aware algorithm for service placement in edge cloud, and adopting a user preference model to learn about user preferences toward services. In addition, \citet{Chiti2019} tackled the problem of virtual function placement in fog-cloud by applying matching game algorithm to fog nodes and virtual functions.

\item[Automation of Cloud Operations]
The scale and the frequent changes of cloud data center networks and configurations necessitate 
faster handling of cloud operations (\ie service management, deployment, and configuration). 
This includes the provisioning of resources and instantiation, configuration, and delivery of services at acceptable service level and without human intervention. 
Thus, orchestration and automation of cloud services have been of high interest to both academia and industry which gave rise to new concepts addressing these concerns. 
For instance, Zero-Touch networking is an emerging concept that leverages SDN and NFV capabilities and aims to automate cloud-based/virtual networks management and operation. 
A related emerging concept is intent-driven networking. It is an end-to-end network configuration automation solution that aims to optimize performance and user experience. It leverages big data and AI technologies to achieve such a goal. 
Preferences and non-functional parameters can be a helpful tool in the design of automated control algorithms. \cite{Sedaghat2013}
\end{description}
\subsection{Limitations and Research Opportunities}
\label{sec:limitations}
\begin{description}
\item[Compactness and Cognitive Burden] As the number of alternatives becomes large, direct assessment of preferences becomes problematic. VMP indeed has huge space of possible alternatives and directly evaluating them would be a demanding cognitive task. For example, consider a customer expressing her preferences for the placement over 10 PMs. She needs to express her preferences in ordinal way through pairwise comparisons of every pair of PMs.  This is indeed problematic as the users would lose focus let alone being cognitively able to accurately specify their preferences.

The vast majority of VMP approaches assume the existence of a utility function. Accurately specifying utility functions by end-users has been shown to be a difficult and tedious task \cite{French}. In many situations, users cannot afford to specify more than qualitative statements of the form ``$x$ is better than $y$". The literature lacks concrete attempts that address user's policies and preferences qualitatively. Furthermore, it is usually assumed that a utility function of one attribute is independent from other attributes. This assumption is too strong to be met in practice as the utility of one attribute may depend on other attributes' values. Generalized Additive Independence Networks (GAI-net) \cite{gonzales2004gai} weaken this assumption by allowing the utility function of one attribute to depend on other attributes. On the qualitative side, Conditional Preference Networks (CP-nets) \cite{1-DBLP:journals/jair/BoutilierBDHP04} are a prominent graphical model to represent conditional preferences.  

Such compact models have been proposed in the last decade to avoid the cognitive burden of specifying preferences in multiattribute domains. They require reasonable effort from the user and adopt the conditional preferential independency notion. This allows for the decomposition of the preferences into smaller but compact components while being expressive enough to capture many scenarios in practice \cite{keeney1993decisions,amor2016graphical}.

\item[Uncertainty, Vagueness, and Risk] Recently, there have been several works on dynamic and uncertain VMP  \cite{lopez2018virtual,chamas2017two,ihara2015many,Mann2015} where workloads and operational statuses may change frequently over time. However, it is important to tackle the problem from the decision maker perspective and consider preferences in uncertain situations. Current approaches usually assume preferences occur with certainty where, in practice, preferences occur with uncertainty over the set of possible alternatives. This is evident in VMP as customers are unaware of possible complexities on the cloud. In such cases, preferences are no longer deterministic choices but rather a  \emph{lottery} over the alternatives. A lottery is a probability distribution over alternatives \cite{keeney1993decisions}. Preferences are then represented over lotteries and known concepts such as expected utility can be used to rank the alternatives.  
\end{description}
 
\begin{description}
\item[Learning and Revision] Machine learning has been recognized as an enabler for building and designing cloud data centers \cite{li2013energy,kundu2012modeling,shi2015mdp}. The VMP problem is dynamic in nature and preferences often change over time. This urges for approaches to not only handle uncertain preferences but also learn to revise them in light of new information. Instead of insisting on having the preferences upfront, it is better to elicit them during the solving process. Such elicitation holds the promise of proposing interactive learning strategies that consider humans in the loop when solving the VMP; thus, lowering the cognitive burden in specifying preferences. Another approach is to learn customers' and administrators' preferences from historical data (\ie learning preferences passively from past deployments).

\item[Group Decision Making] The VMP problem is inherently a group decision making problem. A group of customers and cloud administrators issue their preferences on different aspects of the problem. It is likely that the issued preferences are conflicting with each other and further aggregation and reconciliation methods are required. For instance, consider the scenario of  \citet{Ilkhechi2015} where each group of VMs relies heavily on some PMs (named sinks). These sinks (\eg supercomputers) are a subset of the PMs and it is likely that customers' preferences would be conflicting on how to utilize them.  Thus, the aggregation of preferences needs to ensure certain properties like \emph{fairness} of the deployment among customers. 
\end{description}

\section{Conclusions}
\label{sec:conclusions}
Virtualization technology has enabled cloud data centers to become cost effective, flexible and customizable to host diverse enterprise services and applications at different scales. The VMP is a problem of significant importance to the cloud design for sustaining its profit and popularity. The literature shows a wide variety of approaches to solve the problem and its variants. The problem is also full of preferences of the involved parties including cloud administrators and customers. Therefore, the proper handling of  preferences is crucial to the success of proposed approaches.

In this paper, we discussed the current approaches for handling preferences in the context of VMP. We highlighted certain limitations and identified  research opportunities. Analyzing the VMP problem from a decision making point of view gives rise to several interesting research challenges. Adopting compact models of preferences such as CP-nets and GAI-net is particularly promising given the large number of attributes involved in typical VMP problems. This has the potential of reducing the cognitive overhead of the problem; thus, increase its applicability. Another important research to be explored is interactively learning the constraints and preferences of the cloud admin and/or customers as well as passively learning the soft constraints among VMs and PMs from historical data. Lastly, proposals based on group decision making techniques hold the promise of solving the problem taking everyone's preferences in a fair and transparent manner. Addressing such issues may lead to the development of user-centric VMP that is widely adopted by the stakeholders.

\bibliographystyle{apalike}
\bibliography{vmAzizS.bib}

\end{document}